\documentclass{article} 
\usepackage{lpiabstr_2e}
\usepackage{times}
\usepackage{isolatin1}
\usepackage{graphicx}
\DeclareGraphicsExtensions{.pdf,.jpg,.png,.eps,.ps}
\usepackage[sectionbib]{natbib}

\begin{document}

\title{The Sky in Dust -- Methods and Prospects of Dust Astronomy}

\author{M. Landgraf}
\affil{ESA/ESOC, Robert-Bosch-Strasse 5, 64293 Darmstadt, Germany
(Markus.Landgraf@esa.int)}
\affil{LASP Boulder, CO, USA}
\author{E. Gr\"un}
\affil{HIGP Honolulu, HI, USA}
\affil{MPI-K Heidelberg, Germany}
\author{R. Srama}
\author{S. Helfert}
\author{S. Kempf}
\author{G. Moragas-Klostermeyer}
\author{M. Rachev}
\author{A. Srowig}
\affil{MPI-K Heidelberg, Germany}
\author{S. Auer}
\affil{A\&M Assoc., Basye, USA}
\author{M. Hor\`anyi}
\author{Z. Sternovsky}
\affil{LASP Boulder, CO, USA}
\author{D. Harris}
\affil{HIGP Honolulu, HI, USA}

\runningtitle{THE SKY IN DUST: M. Landgraf et al.}

\titlemake 

\begin{abstracttext}
\section{Introduction}
Information about the make-up of the galaxy arrives in the Solar
system in many forms: photons of different energies, classically
collected by ground- and space-based telescopes, neutral and charged
atomic particles, and solid macroscopic particles: cosmic dust
particles. Dust particles, like photons, carry information from remote
sites in space and time. This information can be analysed in order to
understand the processes and mechanisms that are involved in the
formation and evolution of solid matter in the galaxy.  This approach
is called ``Dust Astronomy'' which is carried out by means of a dust
telescope on a dust observatory in space \citep{gruen05}. The analysis
of cosmic grains collected in the high atmosphere of the Earth has
shown that each dust grain is a small world with various sub-grains
featuring different galactic origin and evolution, which is identified
on the basis of elementary and isotopic analysis
(e.g. \cite{bradley03}). Independent information about the origin and
evolution of the grains coming from the kinematic properties of the
arrival trajectory would be invaluable for linking the isotopic
signature of the formation of heavy elements in old stars and
supernovae to distinctive regions in our galaxy, e.g. known
star-forming regions.

There are different types of dust particles in interplanetary space:
dust from comets and asteroids and interstellar grains traversing the
solar system. The most obvious sources of interplanetary dust are
comets which move on eccentric orbits with a wide distribution of
inclinations with respect to the ecliptic plane \citep{liou99}. The existence of
distinctive low-inclination dust-bands \citep{dermott84} shows that a
significant fraction of meteoroids and dust grains in the zodiacal
cloud have their origin in the asteroid belt.

We know from modelling \citep{liou96b} of the motion of dust in our
solar system that a large mass fraction of it is expelled by close
encounters with the giant planets. Also, the pressure exerted on
sub-micron dust grains by solar radiation makes them leave the Solar
System on hyperbolic trajectories after being released from the parent
body. It can thus be safely assumed that the Solar System is a
prolific source of interstellar dust. In analogy it is clear that
other stars and solar systems form dust clouds around them as they
move through the galaxy. When our Solar System moves through these
clouds these extra-solar dust grains can be collected and analysed. It
was one of the objectives of the Stardust mission \citep{brownlee97} to
collect interstellar dust embedded in the local interstellar cloud
(LIC), which is known to exist from measurements by the Ulysses
mission \citep{gruen93}. Before a sample return of extra-solar dust
from other sources can be attempted, the location and strength of
sources must be identified. Here we propose the Cosmic Dune mission
that is dedicated to characterise the interplanetary and interstellar
dust environment of the Solar System in terms of kinematic,
elementary, and, to a limited extend, isotopic properties.

%\vspace{-5mm}
\section{Dust Flux Distribution}
In order to determine the streams of flux that cross a given point in
the Solar System we can use the information collected by a host of in
situ and ground-based measurements. Based on these measurements a
meteoroid model \citep{dikarev02a} was implemented by the
Max-Planck-Institut für Kernphysik under contract with the European
Space Agency for operational use in the planning of interplanetary
space missions. Here we use the same model in order to determine the
dust flux distribution at a likely location of Cosmic Dune, the
sun ward Lagrange point L$_1$ of the Earth-Sun system.

%explain ESA meteoroid model
The ESA meteoroid model was set up based on data from lunar cratering
record, in situ measurements by the Ulysses, Galileo, and Pioneer
10/11 space probes, and measurements of meteors in the Earth's
atmosphere by radar. In order to give the model predictive power
outside the phase space where the measurements have been taken, the
model invokes the canonical mechanisms of dust production,
transportation, and destruction: mutual collisions of large objects,
Poynting-Robertson drag, and grain-grain collisions. Fore more details
see~\cite{dikarev02a}. The model allows to predict the distribution of
dust streams in the 7-dimensional phase space of mass, space, and
velocity. We assume here that the measurements are performed at L$_1$
and that all grains with masses above $10^{-13}\:{\rm g}$ are
measured. This reduction of the phase space allows us to visualise the
dust flux distribution as a sky-map.

%explain ISD data
The latter population is based mainly on data collected by the Ulysses
spacecraft, which discovered \citep{gruen93} and characterised
\citep{landgraf00a} the properties of the first known interstellar
dust stream.
\begin{figure}[ht]
  \includegraphics[width=\hsize]{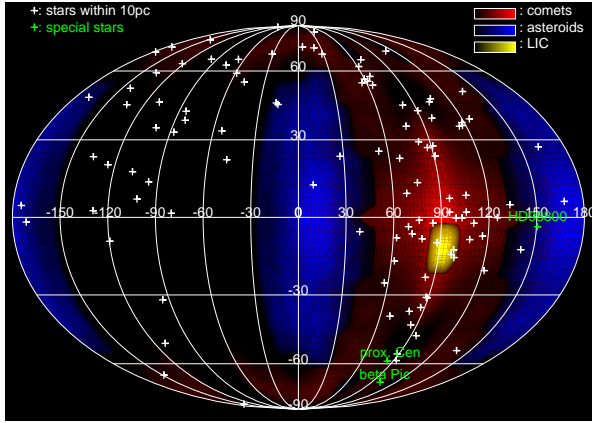}
  \caption{Distribution of dust sources on the sky for a spacecraft
    located at the L$_1$ Lagrangian point at the time of the spring
    equinox. The maximum flux level of grains bigger than
    $10^{-13}\:{\rm g}$ from comets (red) is $70\:{\rm m}^{-2}\:{\rm
      a}^{-1}$, for grains from asteroids (blue) $0.4\:{\rm
      m}^{-2}\:{\rm a}^{-1}$, and from the LIC $600\:{\rm
      m}^{-2}\:{\rm a}^{-1}$.}
  \label{fig:sky} 
\end{figure}

%explain star model
In addition to known dust sources, we can hope to discover new dust
streams crossing the Solar System. It can be speculated that other
solar systems expel dust like ours. Here we consider a much simplified
picture of dust production by nearby stars. It is assumed that the
dust is expelled in a radially symmetrical manner from a dust disk
with thickness $\Delta \phi$ with an average velocity $v$ as a result
of equilibrium dust production by the disk with a constant rate
$\dot{N}$. In this model each star carries along an infinitely
expanding dust cloud through which our Sun moves due to the relative
motion of the two stars, creating a mono-directional dust stream in
our Solar System. The spatial density $n$ of the extra-solar dust
cloud is given by
\begin{eqnarray*}
  n & = & \frac{\dot{N}}{2\pi r^2v \Delta \phi}.
\end{eqnarray*}
Assuming $r=10\:{\rm pc}$, $v=10\:{\rm km}\:{\rm s}^{-1}$, $\Delta
\phi = 20^\circ$, and $\dot{N}=2\times 10^{30}\:{\rm s}^{-1}$ (which
corresponds to an equivalent $10^{-3} M_\oplus$ per year of
$10^{-13}\:{\rm g}$-particles), we arrive at $n\approx 10^{3}\:{\rm
  km}^{-3}$. At a typical relative velocity between stars of $10\:{\rm
  km}\:{\rm s}^{-1}$, this translates to a flux of approximately $300$
particles per square metre and year. Knowing the positions and
relative velocities of nearby stars, we can calculate the upstream
directions of dust streams potentially coming from these
stars. Figure~\ref{fig:sky} illustrates how interplanetary and
interstellar dust sources are distributed over the sky.

\vspace{-5mm}
\section{Instrumentation}
The instrumentation required to perform dust astronomy comprises
devices for measuring the trajectory as well as the elementary and
isotopic composition of cosmic dust grains. Such instrumentation is
currently under development \citep{gruen05} or already available as
flight hardware \citep{srama99}. 
\begin{figure}[h]
  \centering
  \includegraphics[width=\hsize]{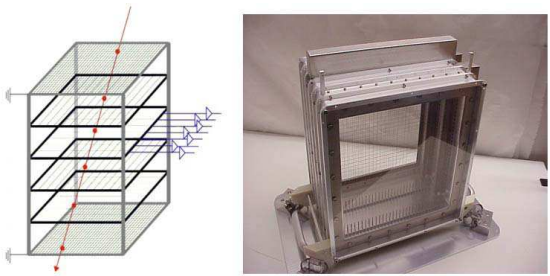}
  \caption{Schematics (left) and hardware (right) of the trajectory
    sensor. The charge signals induced by the passing dust grains are
    amplified and recorded for each of the wires in two pairs of grids
    oriented perpendicular to each other.}
  \label{fig:trajectory_sensor}
\end{figure}
We describe briefly the performance
of a dedicated trajectory sensor and a large-area mass analyser
(LAMA). Figure~\ref{fig:trajectory_sensor} shows the schematics of the
trajectory sensor \citep{gruen05}. Dust particles' trajectories are
determined by the measurement of the electric signals that are induced
when a charged grain flies through a position sensitive electrode
system. The objective of the trajectory sensor is to measure dust
charges in the range $10^{-16}$ to $10^{-13}\:{\rm C}$ and dust speeds
in the range $6$ to $100\:{\rm km}\:{\rm s}^{-1}$. First tests with a
laboratory set-up have been performed and demonstrate the expected
performance. An ASIC charge sensitive amplifier and an ASIC transient
recorder have been developed with a RMS noise of about $1.5\times
10^{-17}\:{\rm C}$.
% trajectory sensor
% LAMA
The principle of time-of-flight mass spectrometry is used to perform
the elementary and isotopic analysis on the dust grains. The
schematics of the instrument shown in figure~\ref{fig:lama_cutaway}
illustrate how ions generated upon impact are accelerated through the
reflectron, where they are measured by a micro-channel plate. The
time-resolved signal from the micro-channel plate produces the atomic
mass spectrum when properly calibrated. The current lab models of LAMA
achieve mass resolutions above 100, which is sufficient to
characterise the elementary and isotopic composition of the most
interesting constituents of cosmic dust grains.
\begin{figure}[ht]
  \centering
  \includegraphics[width=.7\hsize]{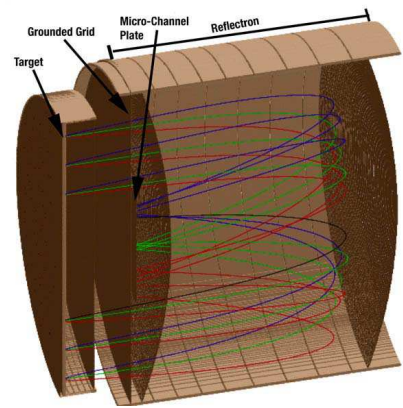}
  \caption{Cut-away schematics of the large-area mass analyser (for
    details see \cite{gruen05}).}
  \label{fig:lama_cutaway}
\end{figure}

%\bibliographystyle{abbrvnat}
%\bibliography{dust}
\end{abstracttext}

\end{document}